\documentstyle[12pt]{article}

\begin{document}

\baselineskip=24pt
\def\beqra{\begin{eqnarray}} \def\eeqra{\end{eqnarray}}
\def\beqast{\begin{eqnarray*}} \def\eeqast{\end{eqnarray*}}
\def\beq{\begin{equation}}	\def\eeq{\end{equation}}
\def\be{\begin{enumerate}}   \def\ee{\end{enumerate}}
\def\fnote#1#2{\begingroup\def\thefootnote{#1}\footnote{#2}\addtocounter
{footnote}{-1}\endgroup}
\def\BM#1{{\boldmath
\mathchoice{\hbox{$\displaystyle#1$}}
           {\hbox{$\textstyle#1$}}
           {\hbox{$\scriptstyle#1$}}
           {\hbox{$\scriptscriptstyle#1$}}}}
\def\haf{\frac{1}{2}}

\begin{center}

{\Large{\bf Quantum Theory and Probabilities}}

\vspace{48pt}

E.C.G. Sudarshan\\
Department of Physics and Center for Particle Physics\\
University of Texas, Austin, Texas 78712-1081\fnote{*}{e-mail: 
sudarshan@physics.utexas.edu}
\end{center} 

\begin{abstract}{It is often stated that quantum mechanics only makes statistical predictions and that a
quantum state is  described by the various probability distributions associated with it.  Can we describe a
quantum state completely in terms of probabilities and then use it to describe quantum dynamics?  What is the
origin of the probability distribution for a maximally specified quantum state?  Is quantum mechanics `local'
or is there an essential nonlocality (nonseparability) inherent in quantum mechanics?  These questions are
discussed in this paper.  The decay of an unstable quantum state and the time dependence of a minimum
uncertainty states for future times as well as past times are also discussed.}
\end{abstract}

\vfil
\pagebreak

\section{Introduction: Classical Dynamics}

The standard form of classical dynamics involved first-order equations of motion of the phase point 
representing the system.  For a collection of particles the phase point is specified by 6N variables, 3N
position coordinatres and 3N momenta.  The equations of motion would then furnish the trajectory of this
phase point.  These are nonintersecting continuous curves (except at the points where the Hamiltonian is
irregular).  Given an initial phase point, the dynamics specifies the final point for any time.  `Adjacent'
phase points often lead to adjacent end points.  (Note that in a symplectic space like the phase space, there
is no canonically defined `distance'.) 

The `atomic' propositions for a classical system are phase points; in a state
specified by an atomic proposition all dynamical variables have definite values.  All dynamical variables are
compatible observables which can be simultaneously specified.[1] 

In many cases one may choose a non-external
initial condition: given by a (nonnegative) phase-space density.  Then for a Hamiltonian evolution this goes
into a statistical state.  More generally, the statistical state is a distribution on the phase space which
maps dynamical variables into numbers.  Only distributions which are `measures' in that they are real and
nonnegative everywhere are allowed.  A simple alternate characterization is by the expectation value

\begin{equation}
\chi(\lambda,\mu)=  \langle\exp\,i(\lambda p+\mu q)\rangle
\end{equation}
which is called the characteristic function.[2]  It is bounded in magnitude by unity.

In addition to the distribution nature of the initial condition, we have two other cases where it is useful 
to consider such statistical descriptions.  One is for a nonlinear system which has very complicated
trajectories which are unstable: for small changes in the initial condition the final state is vastly
different.  In these cases the trajectory description is not useful,[3] even though the trajectories do not
cross each other (except for phase points which are singular). 

Another context is for stochastic dynamics
which do not map points into points but only distributions into distributions.  A natural origin for such a
behavior is to have an open dynamical system.  The stochastic behavior is classified into Markovian and
non-Markovian processes [4]: in the Markovian case only the immediate past is sufficient to determine the
future evolution; it has no `memory'.  A non-Markovian process can be `lifted' to be a Markovian system with
a larger phase space.  [Example: Multiple scattering of a changed particle: transverse displacements alone do
not give a Markovian process, but if we include the slopes of the trajectories also, we can get a Markovian
process.][5]

One method of getting a stochastic process from a deterministic dynamics is by `contraction', that
is ignoring some degrees of freedom.  In many cases we can `lift' the stochastic process into a deterministic
dynamics of an extended dynamical system.  

A stationary Markovian stochastic process possesses the semigroup
structure: if $A(t)\rho$   is the map of the phase space distribution $\rho$  by the Markovian
process, then 
\begin{equation}
A(t_1)A(t_2)\rho=A(t_1+t_2)\rho\;, ~~t_1>0,t_2>0\;.
\end{equation}
The map of the distribution is an `into' map corresponding to a contraction.  The contractive semigroup has
a  dissipative part.  The generator of the semigroup has an `imaginary' part. 

It is often stated that for a
deterministic dynamical evolution with very complicated trajectories, there is a loss of information in the
forward direction and hence the evolution produces an arrow of time.[6]  This `fundamental resolution', first
offered by Boltzmann and then refined by a succession of authors, raises a question: how can a reversible
evolution lead, by itself and without approximation, to an irreversible process?  But if we trace the
evolution to the past from the present state, we find that there is again an apparent `loss of information'
to the past!  In other words, there is an unstable trajectory which leads to the simple present state only
appears to have lost information.  So within the closed dyamical system with a reversible dynamics, an arrow
of time cannot be discerned.  This is essentially the Loschmidt objection to Boltzmann's assertion.   The
unstable, nonlinear system makes it less obvious but the Loschmidt dejection still obtains.  It is also
important to note that only `reversibility' is sufficient: `time reversal invariance' is not required.

\section{Quantum Dynamics: Characteristic Functions and \\ Distributions}

Quantum systems are characterized by the superposition principle [7]:  atomic (external) states can be
obtained  which are superpositions of two distinct states.  This property automatically leads to
noncommuting dynamical variables which cannot be simultaneously  specified.  For example, a wave function
$\psi({\BM x})$   yields   a position probability distribution
\begin{equation}
P({\BM x})=|\psi({\BM x})|^2 
\end{equation}
so that
\begin{equation}
f(x)\rightarrow\langle f(x)\rangle = \int\,P(x)\,f(x)dx\;.
\end{equation}
The position probability distribution has less information than $\psi({\BM x})$  
 (even $\psi({\BM x})$         module an overall phase) since there is no information about
the momentum: but given $\psi({\BM x})$           we can get the momentum space function

\beq
\tilde{\psi}({\BM p})=\int\, (2\pi)^3  \int\psi (x)e^{ipx}\;dx\,.
\eeq
from which the  momentum-related dynamical variables have the expectation value
\beq
g({\BM p})\rightarrow\langle g({\BM p})\rangle=\int\,\tilde\psi({\BM p})\psi(p)\, g(p)dp\,.
\eeq
But, what about functions of both $p$ and $q$?  For the characteristic function
\beq
\chi(\lambda,\mu)=\langle e^{i\lambda p+i\mu q}\rangle
\eeq
we may make a canonical transformation
\beq
q\rightarrow\frac{\lambda P+\mu Q}{(\lambda^2 +\mu^2)^\haf}\;,\;p\rightarrow \frac{(-\mu P+\lambda
Q)}{(\mu^2+\lambda^2)^\haf}\,.
\eeq
Then
\beq
\chi(\lambda,\mu) = \langle e^{iQ}\rangle =\int\tilde\Psi(Q)
e^{iQ\sqrt{(\lambda^2+\mu^2)}}\,\psi(Q)dQ\;.
\eeq
Since any bounded operator of $q$ and $p$ can be expressed in terms of the unitary operators
$\exp\{i(\lambda p+\mu q)\}$, their expectation values can be computed given the wave function. 

The
characteristic function can be computed in terms of the wave function $\psi(q)$    as follows:
\beq  
\exp(i(\BM{ \lambda p}+\BM{ \mu q}) = \exp\left(i\frac{\BM{\lambda p}}{2}\right) \exp(i{\BM{ \mu q}})\; 
\exp\left(i\frac{\BM{\lambda p}}{2}\right)\;.
\eeq
Hence
\beq
\langle\exp(i\lambda^*(\BM{xp}+\BM{\mu q}) \rangle  =\int\psi^*(\BM{x})\, e^{i\BM{xp}/2}\;e^{i\BM{\mu q}}\;
 e^{ixp/2}(\BM{ x} d\BM{ x})= \int\psi^*\left(\BM{x}- \frac{\lambda}{2}\right)\,e^{i\mu x}\,\psi\left(\BM{x}+
\frac{\lambda}{2}\right)d\BM{ x}\;.
\eeq

These calculations pertain to the atomic propositions corresponding to `pure states.'  For a
statistical state  we take mixtures of states with suitable probability weights according to
\beqra
\rho_1 &=& \psi_1\psi_1^\dag~,~\rho_2=\Psi_2\psi_2^\dag \nonumber\\
\rho &=& \cos^2\theta\,\rho_1 + \sin^2\theta\,\rho_2
\eeqra
and more generally
\beq
\rho=\Sigma p_i\rho_i ~,~\Sigma p_i=1~,~p_i\leq 0\,.
\eeq
The characteristic functions obey the same law:
\beq 
\chi\,(\BM{\lambda,\mu})=\Sigma p_i\lambda_i (\BM{\lambda,\mu})\;,\;\Sigma p_i=1,\;p_i\geq 0\;.
\eeq

For classical dynamics, given the characteristic function $\chi({\lambda,\mu})$  , the
multivariate  phase-space distribution function is given by the double Fourier transform (for $N$ particles):
\beq(
\rho({\BM p,q}) = 2\pi)^{-bN} \int\!\!\int e^{-i(\BM{\lambda,p} +\BM{\mu q}})\chi
(\BM{\lambda,\mu}) d\BM{\lambda}d\BM{\mu}\;.
\eeq
 By definition the phase space density $\rho(\BM p,\BM q)$  is nonnegative and integrate
to unity:
\beq
\rho({\BM p,q})\geq 0~~,~~\int\!\!\int\rho({\BM p,q})\; pdq=1\;.
\eeq
So $\rho$  is a proper probability measure.  
The same construction may be carried out for quantum characteristic functions:[8]
\beq
W ({\BM p,q})\,\frac{1}{(2\pi)^{6n}}\int\!\!\int\chi(\BM{\lambda,\mu})d\BM\lambda d\BM\mu\;.
\eeq
The normalization condition is still valid.
\beq
\int\!\!\int W(\BM{p,q}) dpdq=1\;.
\eeq
But it is no longer nonnegative.  $W$ may become negative pointwise in the phase space.  There is however a 
positivity condition:
\beq
\langle(f(\BM{p,q}))^2\rangle~~\geq 0\;.
\eeq
$W$ is the Wigner phase space density.[9]  It also satisfies the condition
\beq
\int\!\!\int\;W^2(\BM{p,q})d\BM p d\BM q =\frac{1}{(2\pi)^{3N}} = tr\rho
\eeq
for a pure state and
\beq
\int\!\!\int\;W^2(\BM{p,q}) d\BM{ p} d \BM q =\frac{tr(\rho^2)}{(2\pi)^{3N}}< 1\,.
\eeq

Given the phase space density $\rho(\BM{p,q})$           or $W(\BM{p,q})$, the equations of
motion can be transcribed in  terms of these.  For classical dynamics we have the Liouville equations
\beq
\frac{d}{dt}\,\rho(q,p,t)=-i[H,\rho](q,p,t)\;.
\eeq
while for quantum dynamics[10]
\beq
\frac{d}{dt}\,W(q,p,t)=(M\,W) \;(q,p,t)
\eeq
where the Moyal operator $M$ is defined as 
\beq
M\rho(pq)=\frac{2}{\hbar}\, H(p,q)\,\left\{\sin\frac{\hbar}{2}
\left( \frac{\stackrel{\leftarrow}{\partial}}{\partial q}
 \,\frac{\stackrel{\rightarrow}{\partial}}{\partial p} -
\frac{\stackrel{\leftarrow}{\partial}}{\partial p} \frac{\stackrel{\rightarrow}{\partial}}{\partial
q}\right)\right\}\rho(p,q)\,. 
\eeq

This is a nonlocal operator (except for a polynomial Hamiltonian) but it preserves the purity of the state.
In place of canonical variables we may have other variables which do not commute.  A particularly important 
case is that of a spin system either by itself or in terms of spinning particles.  We will discuss this later
in this presentation.

\section{Distributions over Non-commuting Variables: Spectra}

If $B,C$,   are noncommuting operators for a quantum system, then with every dynamical variable we
have  spectrum of possible `eigenvalues;' and according to the von Neumann postulate, any measurement yields
one of these eigenvalues with a frequency of each eigenvalue depending on the state $\psi$  (or $\rho$ )
of the quantum system.  Every state assigns for any dynamical variable $B$ a probability distribution, and
the expectation value of any function of $B$ can be computed using this probability distribution.  If $B$
and $C$ commute, then we can consider a spectrum for the pair ($B, C$); and so on.   For example, for a
spinning particle in an inhomogeneous magnetic field, we can specify both the momentum components and the
spin projection along the gradient of the magnetic field.  The first has a continuous spectrum and the
second a discrete Stern-Gerlach spectrum. The measurement postulate[11] leaves out the definition of
`measurement'.  For Bohr and Heisenberg this `quantum jump' into an eigenvalue is instantaneous and
indescribable; but Schr\"odinger insists that it is a process however rapid the transition [12].  We will
comment on this later. Given a quantum state and an operator $B$, we have a spectrum of values.  What we can
calculate is the probability distribution but not the actual value obtained in an experiment.  In this
respect, quantum probabilities are no different from classical probabilities: any particular `realization'
is random.  What we can compute or predict is the probability distribution.  In classical physics we do not
appeal to the `collapse of a probability distribution' into a unique measured value: why should we then talk
about the collapse of a quantum distribution?  In both cases an immediate remeasurement yields the same
definite value.  The only difference between the observed and quantum distribution is that in the former we
can have a probability distribution for the dynamical variable, but in the latter we can specify at most a
complete commuting  set.  In particular, for the classical case, since    the equation of motion change a
dynamical variable into a dynamical variable, we can have a complete history of the particle's time
evolution.  For a Hamiltonian system these are a set of continuous nonintersecting curves.  For more general
stochastic evolution the histories can repeatedly branch out into the future.

\section{Spin Distributions: Separability}

For a quantum system we must remember that not all dynamical variables can be simultaneously measured.  At 
any one time a complete commuting set can be measured and this measurement yields a collection of spectral
values.  What about the evolution in time?  At  a later time this set can be measured; thus we can construct
a set of histories for a given state; but for this to be consistent we must impose some consistency
conditions.  This becomes necessary since states can superpose: so instead of continued branching we have
also interfering recombinations.  If we can assign probabilities for all measurements for several times, we
talk of a consistent history.[13] 

Any history in which there is no recombination is automatically consistent.  The most
general consistent history is when there is at most one loop on any set of histories: (the $\pi$'s  are
projections)
\beq
|\psi_i^\dag\,\pi_1\,\pi_2\,\pi_3\ldots\psi_j|^2 = P_{ij}(\pi_1\pi_2\pi_3\ldots)\,.
\eeq
If these are to be consistent, then
\beq
P_{ij}(\pi_1\;\pi_2\ldots) =0 \cdot i\neq j;.
\eeq
If the $\pi$'s  are one-dimensional projections, we can have the pairs of interfering
amplitudes be out of  phase by 90 degrees and hence provide no interference in probability.[14] 

While
consistent histories require nonnegative probabilities, we can raise the question of the characteristic
function providing a phase space distribution.  For canonical variables $p,q,$ this is given by the Wigner
function.  When we want to find the probability distribution for a commuting set, namely functions of 
\beq
x=\lambda p+\mu q-\nu
\eeq 
then we get a nonnegative-time probability distribution:
\beq
P(\lambda,\mu,\nu) =\int\!\!\int W(p,q)\;\delta(\lambda p+\mu q-\nu)\,dpdq\;.
\eeq
This is a nonnegative distribution.  $W(q,p)$   determines the quantum tomogram  $P(\lambda,\mu,\nu)$; and
if  the $P(\lambda,\mu,\nu)$   is known for all $\lambda,\mu,\nu$, then it completely
determines $W(p,q)$. Note that the tomograms scale according to 
\beq
P(\lambda,\mu\nu)=\frac{\nu_0}{\nu}\; P(\lambda\nu_0,\,\mu/\nu_0,\, \nu/\nu_0).
\eeq
so that it is really a function of only two variables.

In the more general case of any quantum system we could compute a `master distribution function' from which 
the marginal distributions for any subset of variables can be computed by integrating over all the unwanted
variables.   This is particularly useful in computing probability distributions for spin projections along
various directions.  No two of these commute so that the master distribution is not positive definite.  For
example, given a spin-1/2 object the distribution in a state $\rho$    for $\BM{ s\cdot n}_1$, and
$\BM{ s\cdot n}_2$ is given by 
\beq
 P(n_1,n_2) =\langle (s\cdot n_1)(s\cdot n_2)\rangle = \psi^\dag \frac{1+s\cdot n_1}{2}\;
\frac{1+s\cdot n_2}{2}\; \psi=tr(\rho\pi(n_1)\,\pi(n_2))\;.
\eeq
This quantity is not real, but its marginal distributions are nonnegative.
\beq
P(n_1)=\sum^1_{+,-}\,P(n_1,\pm)=tr(\rho\pi(n_1)) \geq 0\,.
\eeq
Instead of giving the state $\rho$ if we use the projection $\frac{1+S\cdot n}{2}$ for $\rho$     we
get
\beq
P(n,n_1,n_2) = tr \left(\frac{1+S\cdot n}{2}\; \frac{1+S\cdot n_1}{2}\;\frac{1+S\cdot n_2}{2}\right)
\eeq
which is in general complex but gives the correct marginal distributions.  In place of $P(n,n_1,n_2)$  we
could  take the symmetrized quantity 
\beq
P_s(n,n_1,n_2) = {\rm symmetrized}~ P(n,n_1n_2)
\eeq 
which is real but not positive definite.  These also give the correct (symmetrized) marginal distributions.

These observations give us the proper tools for dealing with Bell's inequalities.  If we consider the master 
distributions for three directions $n_1,n_2,n_3$ , then
\beq
P(n_1,n_2) = \sum_{n_3=\pm} P(n_1,n_2,n_3)\;{\rm etc.}
\eeq
If $P(n_1n_2n_3)$      were a classical nonnegative probability distribution, then these two-point
correlations satisfy  the triangle inequalities
\beq
P(n_1,n_2)+P(n_1,n_3)\geq P(n_2,n_3) \;{\rm etc.}
\eeq
However, quantum mechanics gives the unsymmetrized expression
\beq
P(n_1n_2n_3) = tr\left\{ (1+s\cdot n_1) (1+s\cdot n_2) (1+s\cdot n_3)\right\} 
\eeq 
which on symmetrization yields
\beq
\frac{1}{4}(1+\cos\theta_{12}+\cos\theta_{23}+\cos \theta_{31})\,.
\eeq

If we suitably choose $\theta_{12},\theta_{23},\theta_{31},$ we can easily violate the triangle
inequalities and hence Bell's  inequalities [16]. 

In the vast literature on Bell's inequalities many authors
ascribe the violation of Bell's inequalities to the essential nonlocality (nonseparability) of quantum
mechanics!  But we see that if we accept indefinite master distributions, there is no need to invoke
nonlocal properties, least of all to nonlocal hidden variables.

\section{Does Reversible Dynamics Furnish an Arrow of Time?}

We now turn our attention to another fundamental problem: How to obtain irreversibility and an arrow of time 
from a closed dynamical system with time-reversible equations of motion?  We have indicated already why the
demonstrations of an arrow may be an unjustified conclusion.  Note that any approximation in calculating a
unitary operator would lead to nonunitarity and a dissipative evolution.  But such a `mistake' is not a
demonstration of the arrow of time!  If the system is open   the forward evolution is made stochastic by
keeping the interaction with the external system; since we use this only for the forward evolution but not
the backward evolution, we have put in an arrow of time `by hand'.

\section{Decaying States in Quantum Mechanics}

The most notable case of this kind is the decay of a metastable state (or an unstable particle).  The quantum 
theory of spontaneous deexcitation of an excited atom was formualted by Dirac. [17]  Under the influence of
the perturbative radiation field coupling , the excited state goes into a superposition of the excited state
and a continuum of photons plus the ground state.  This is a unitary transformation and phase relations are
preserved.  If we develop the state backwards in time, we get to the pure excited state; and even further
back a coherent combination of the excited state with the ground state.  No irreversibility and no arrow of
time obtains at this stage of computations.  However, if we ask what is the probability of the survival of
the unexcited state, then we get what appears to be a decay: but that is no more than the `decay' of the
magnitude of the square of the x-component of a vector rotating about the origin.  Only thing to be specially
invoked is the continuum of photon frequencies and the more or less montonic   decrease in
the probability of the survival of the excited state.  By deliberately ignoring the phase relations we obtain
a probability distribution.  It may be approximated by a dissipative stochastic process, dominated by an
exponential decay; and it appears to provide an irreversible process.  But if we had traced it back, we would
get an antidissipative process.  There is no arrow unless we deliberately ignore the negative time
propagation.

\section{Analytic Continuation of the Resolvent: Dual Space Formalism}

The problem can be treated in a more precise fashion by looking for the exact time development using the 
resolvent of the total Hamiltonian.  This resolvent can be analytically continued into the complex plane 
without      changing the time development [18].   The complex spectrum of the resolvent may be
identified with the analytic continuations of the self-dual Hilbert space into a pair of dual vector spaces. 
Depending on whether we are interested in the forward time development or the backward time development, the
analytically continued spectrum is most conveniently chosen to be in the lower or the upper half plane.  Both
evolutions are irreversible; but if we make approximations, we get semigroups which are not automatically
reversible.  The irreversibility is not in the system but the mutilation of the time development.

\section{Concluding Remarks: Quantum Radiative Transfer}

This reversibility paradox is well illustrated by the time development of a minimum uncertainty wave packet.  
It appears that the wave packet spreads in the future but it also spreads in the past.  The increase in width
depends on the elapsed time squared. The propagation of partially coherent light may be useful to illustrate
the difference between the unitary evolution of the amplitude and the `irreversible' spreading of the
intensity.  Classical radiative transfer uses the spectral intensity distribution.  The classical radiative
transfer theory[19] is not sufficient when we have interference, like passage of light through a double slit
or through a diffraction grating.  We do not have consistent histories since interference dominates.  So the
language used in `delayed choice' and other gedanken experiments is somewhat inappropriate.  But we can have
a generalized radiative transfer formulation in which indefinite spectral and angular distribution are
used  [20].  The Wolf function, which is the analogue of the Wigner function generalized to a field, would
then be the object of the generalized radiative transfer and it could treat all light propagators.  If we
include polarization also, we need polarization 2x2 matrices in place of specific intensities with angular
dependence.  We hope to present this in a subsequent paper.

\noindent
 {\Large{\bf References}}
 
\begin{enumerate}
\item  See, for example, G. W. Mackey, ``Mathematical Foundations of Quantum Mechanics; 
a lecture-note volume", New York, W. A. Benjamin (1963)\\  
I. E. Segal, ``Postulates for General Quantum Mechanics", Annals of Mathematics,
(2) {\bf 48}, 930 (1974)\\
 E. C. G. Sudarshan, ``Structure of Dynamical Theories",
Brandeis Lecture Notes, W. A. Benjamin (1960).
\item H. Cramer, ``Mathematical Methods of Statistics", Princeton University Press
(1946).
\item  V. I. Arnol'd and A. Avez, ``Ergodic Problems of Classical Mechanics", New York,
 Benjamin (1968) and V. I. Arnol'd, ``Mathematical Methods of Classical
Mechanics", New York: Springer-Verlage (1978)\\
 I. Prigogine, ``The End of
Certainty: Time, Chaos, and the New Laws of Nature", Free Press, New York (1997)\\
D. J.  Dreibe, ``Fully Chaotic Maps and Broken Time Symmetries",  Kluwer Academic
Pub., Boston (1999).  See however, ~~~~~~~ Huw, ``The Arroco of Time" (Oxford~~~~~~).
\item  I. I. Gikhman, ``The Theory of Stochastic Processes", Springer-Verlag, New York (1974).
\item  L. Eyges, ``Multiple Scattering with Energy Loss," Phys. Rev. {\bf 74}, 1534 (1948)\\  
S. Biswas, E. C. George, B. Peters and M. S. Swamy,  ``Mass Determination on Steeply Dipping 
Tracks in Emulsion Block Detectors", Nuovo Cimento Supplemento {\bf Ser. 9, v.
12}, 369 (1954)\\ P. J. Lavakare and E. C. G. Sudarshan,   ``The Study of Spurious
Scattering in Nuclear Emulsions and the Effect of Higher Order Differences in
Scattering Measurements", Nuovo Cimento Supplemento {\bf Ser. 10, v. 26}, 251
(1962)
\item  T. Petrosky and I. Prigogine, ``The Liouville Space Extension of Quantum Mechanics", 
Advances in Chemical Physics, Vol. XCIX, John Wiley, New York (1997)
\item  P.~A.~M. Dirac, ``Principles of Quantum Mechanics" ( 4th Edition), Clarendon Press, 
Oxford (1981)
\item  See,  for example, C. L. Mehta and E. C. G. Sudarshan, ``Relations between Quantum 
and Semiclassical Description of Optical Coherence", Phys. Rev.  {\bf 138}, B274
(1965)
\item  E. P. Wigner, ``On the Quantum Corrections for Thermodynamic Equilibrium", Phys. Rev. 
{\bf 40}, 749 (1932)
\item  J. E.  Moyal, ``Quantum Mechanics as a Statistical Theory", Proc. Camb. Phil. Soc. 45, 
99 (1949)
\item J. von Neumann, ``Mathematical Foundations of Quantum Mechancis",
Princeton University Press (1955).
\item See the discussion in I.~M. Duck and E.~C.~G. Sudarshan, ``100 Years
of Planck's Quantum", World Scientific, Singapore (2000).
\item M. Gell-Mann and J.~B. Hartle, ``Alternative Decohering Histories in Quantum
Mechanics", Proc. of the 25th International Conference on High Energy Physics,
Vol. 2, World Scientific, Singapore (1991), pp.  1303-10\\
R. Omnes, ``General Theory of the Decoherence Effect in Quantum Mechanics", Phys.
Rev. {\bf A56}, 3383 (1997) and R. Omnes, ``Consistent Interpretations of Quantum
Mechanics", Reviews of Modern Physics {\bf 64}, 339 (1992)\\
R. B. Griffiths, ``Consistent Histories and the Interpretation of Quantum
Mechanics", J. Stat. Phys. {\bf 36}, 219 (1984).
\item  E. Chisom, T.~F. Jordan and E.~C.~G. Sudarshan, ``Weak Decoherence and
Quantum Trajectory Graphs", Int. J. Theor. Phys. {\bf 35}, 485 (1996).
\item  V.~I. Manko, ``Symplectic Tomography as Classical Approach to Quantum
Physics", Phys. Lett. {\bf A213}, 1 (1996).
\item  T. Rothman and E.~C.~G. Sudarshan, ``A New Interpretation of Bell's
Inequalities", Int. J. Theor. Phys. {\bf 32}, 1077 (1993). 
\item  W. Heitler, ``The Quantum Theory of Radiation", Clarendon Press, Oxford
(1963).
\item  E.~C.~G. Sudarshan, C.~B. Chiu and V. Gorini, ``Decaying States as
Complex Energy Eigenvectoers in Generalized Quantum Mechanics", Phys. Rev. {\bf
D18}, 2914 (1978)\\
E.~C.~G. Sudarshan and C.~B. Chiu,``Analytic Continuation of Quantum
Systems and Their Temporal Evolution", Phys. Rev. {\bf D47}, 2602 (1993)\\
E.~C.~G. Sudarshan, ``Quantum Dynamics, Metastable States, and Contractive
Semigroups", Phys. Rev. {\bf A46}, 37 (1992)\\
 E.~C.~G. Sudarshan, ``Quantum Dynamics in Dual Spaces", Phys. Rev.{\bf A50},
2006 (1996).
\item  S. Chandrasekhar, ``Radiative Transfer", Dover Publications, New York
(1960).
\item E.~C.~G. Sudarshan, ``Quantum Theory of Radiative Transfer", Phys. Rev. {\bf
A23}, 2802 (1981).

\item E.~C.~G. Sudarshan, ``Quantum Electrodynamics and Light Rays", Physica A 
{\bf 96A}, 315 (1979).

\end{enumerate}
\end{document}